\documentstyle[prd,aps]{revtex}

\begin{document} 
\draft

\title{Time Machines Constructed from Anti-de~Sitter Space}
\author{Li-Xin Li}
\address{Department of
Astrophysical Sciences, Princeton University, Princeton, NJ 08544}
\date{November 12, 1998}
\maketitle

\begin{abstract}
In this paper time machines are constructed from anti-de~Sitter space.
One is constructed by identifying points related via boost transformations
in the covering space of anti-de~Sitter space and it is shown that
this Misner-like anti-de~Sitter space is just the Lorentzian section of
the complex space constructed by Li, Xu, and Liu in 1993. The others are
constructed by gluing an anti-de~Sitter space to a de~Sitter space, which
could describe an anti-de~Sitter
phase bubble living in a de~Sitter phase universe. Self-consistent vacua for 
a massless conformally
coupled scalar field are found for these time machines, 
whose renormalized stress-energy tensors are
finite and solve the semi-classical Einstein equations. The 
extensions to electromagnetic fields and massless neutrinos are discussed. 
It is argued that, in order to make the results consistent with Euclidean 
quantization,
a new renormalization procedure for quantum fields in Misner-type spaces
(Misner space, Misner-like de~Sitter space, and Misner-like anti-de~Sitter 
space) is required. Such a ``self-consistent" renormalization procedure is
proposed. With this renormalization procedure, self-consistent vacua exist
for massless conformally coupling scalar fields,
electromagnetic fields, and massless neutrinos in these Misner-type spaces. 
\end{abstract}


\section{Introduction}
In classical general relativity there are many solutions of Einstein equations
with closed timelike curves (CTCs) 
\cite{god49,tau51,new63,mis67,mor88,got91}. However, 
some early calculations of vacuum
polarization in spacetimes with CTCs indicated that the renormalized 
stress-energy tensor diverged at the Cauchy horizon or the polarized 
hypersurfaces \cite{his82,fro91,kim91,gra93}. Hawking thus proposed
the chronology protection conjecture which stated that physical
laws do not allow the appearance of CTCs \cite{haw92}. But, many 
counter-examples to the chronology protection conjecture have been found
\cite{bou92,li93,li94,tan95,li96,kra96,sus97,vis97,cas97,li98,got98}.
In particular, Li and Gott \cite{li98} have found a self-consistent vacuum
for a massless conformally coupled scalar field in Misner space
(see also \cite{cas97}), which 
gives an example of a time machine (i.e. a spacetime with CTCs)
at the semi-classical level (i.e.
the background spacetime is classical but the matter fields are quantized).
 
Of more interest, recently Gott and Li have discovered that 
CTCs could play an important role in the early universe: if we trace backward 
the history of time, we may enter an early epoch of CTCs, which means that 
there is no earliest event in time \cite{got98}. According to the theory of 
quantum foam 
\cite{whe55}, in the early universe (at the Planck epoch), quantum fluctuations
of spacetime should be very important and the spacetime might have 
a very complicated topology. Very strong
fluctuations in the metric of spacetime could cause the lightcones
to distribute randomly, which could give rise to a sea of CTCs in
the early universe. 
Therefore, we might expect that at very early epochs,
the universe could have a tangled network of CTCs. 

One model of the creation of the universe is the model of ``tunneling 
from nothing"
\cite{vil82,har83}. In this model the universe is supposed to be a 
Lorentzian spacetime [with signature $(-,+,+,+)$] glued to an 
early Euclidean space
[with signature $(+,+,+,+)$], thus the universe 
has a beginning in time (i.e. the beginning of the Lorentzian section). This
model of ``tunneling from nothing" has some shortcomings 
(see \cite{got98}, and Penrose in \cite{haw96}). Contrasting with the model of 
``tunneling from nothing", in the model of Gott and Li \cite{got98}, 
the universe
does not need a signature change and has no beginning in time. 
Gott and Li's universe is always a Lorentzian spacetime but at 
a very early epoch
there is a loop of time.  
The universe could thus be its own mother and create itself. 
The model of Gott and Li has some additional
interesting features: the present epoch of
the universe is separated from the early CTCs epoch by a past chronology
horizon. The only self-consistent solution with this geometry has pure
retarded potentials, creating naturally  
an arrow of time in our current universe, which is 
consistent with our experience \cite{got98}. Thus, CTCs have potentially
important applications in the early universe.

Anti-de~Sitter space is a spacetime which has CTCs everywhere. It
is a solution of the vacuum Einstein equation with a negative cosmological
constant and has maximum symmetry \cite{haw73}. Anti-de~Sitter space plays
a very important role in theories of supergravity and superstrings 
\cite{cas91,wit98}. 
If we ``unfold" anti-de~Sitter space and go to
its covering space, the CTCs disappear. However, if we identify the events
related by boost transformations in the covering space of anti-de~Sitter
space, we will get a spacetime with an infinite number of regions with 
CTCs and an infinite number of regions without CTCs, where the regions with CTCs
and the regions without CTCs are separated by chronology horizons. 
The causal structure is similar to that of Misner space, except that
Misner space has only two regions with CTCs and two regions without
CTCs. For Misner space, Li and Gott have found a self-consistent vacuum for
a massless conformally coupled scalar field, which is an ``adapted'' 
Rindler
vacuum (i.e. a Rindler vacuum with multiple images) \cite{li98}. In this
paper we will show that a self-consistent vacuum also exists for a
massless conformally coupled scalar field in the Misner-like 
anti-de~Sitter space constructed above, which is simply the conformal
transformation of Li and Gott's adapted Rindler vacuum.

In 1993, Li, Xu, and Liu constructed a time machine in a space with
a complex metric
\cite{li93}. In this paper we will show that the Lorentzian section of 
that solution is just the Misner-like anti-de~Sitter space described
above.

Inflation theory proposes that during an early epoch the universe was in a
state with an effective positive cosmological constant at the GUT (or Planck)
scale \cite{gut81,lin82,alb82}, which is well-described as
a de~Sitter phase. By transition to a zero cosmological
constant (through either quantum tunneling or classical evolution), 
the universe then enters a Friedmann Big Bang stage. 
But, if there is a transition to
a negative cosmological constant (it does not seem physical theories 
exclude a negative cosmological constant --- especially since in supergravity
and superstring theories anti-de~Sitter space is the only 
known self-consistent solution besides Minkowski space
\cite{cas91}), the universe could enter an anti-de~Sitter phase
where CTCs exist. In this paper we will consider some
models describing the transition 
between a de~Sitter space and an anti-de~Sitter space, 
which are obtained by gluing a de~Sitter space to 
an anti-de~Sitter space along a bubble wall, and we will show that 
self-consistent vacua for these solutions also exist. 

The generalization to the case of
electromagnetic fields and massless neutrinos
will also be considered. It is argued that, in order to be
consistent with Euclidean 
quantization,
a new renormalization procedure for quantum fields in Misner-type spaces
is required. A ``self-consistent" renormalization procedure is then
proposed. With this renormalization procedure, self-consistent vacua exist
for massless conformally coupled scalar fields,
electromagnetic fields, and massless neutrinos in the Misner-type spaces. 

\section{Misner-like Anti-de~Sitter Space}
\label{sec2}
Anti-de~Sitter space $AdS^4$ is a hyperbola
\begin{eqnarray}
   V^2+W^2-X^2-Y^2-Z^2=\alpha^2
   \label{ades}
\end{eqnarray}
embedded in a five-dimensional space $R^5$ with metric
\begin{eqnarray}
   ds^2=-dV^2-dW^2+dX^2+dY^2+dZ^2.
   \label{metr0}
\end{eqnarray}
Anti-de~Sitter space has topology $S^1\times R^3$ and 
is a solution of the vacuum Einstein equations with
a negative cosmological constant $\Lambda=-3/\alpha^2$, which has
maximum symmetry (i.e. it has ten Killing vectors).
 
Several coordinate systems can be defined in anti-de~Sitter space:
1) {\em Global static coordinates}. Define
\begin{eqnarray}
   \left\{\begin{array}{l}
   V=-\alpha\cosh\chi\cos{t\over\alpha}\\
   W=\alpha\cosh\chi\sin{t\over\alpha}\\
   X=\alpha\sinh\chi\cos\theta\\ 
   Y=\alpha\sinh\chi\sin\theta\cos\phi\\
   Z=\alpha\sinh\chi\sin\theta\sin\phi
   \end{array}\right.,
   \label{coord1}
\end{eqnarray}
where $0\leq t/\alpha< 2\pi$, $0<\chi<\infty$, $0<\theta<\pi$, and
$0\leq\phi<2\pi$. 
Then the anti-de~Sitter metric can be written as
\begin{eqnarray}
   ds^2=-\cosh^2\chi ~dt^2+\alpha^2 d\chi^2+\alpha^2\sinh^2\chi
   \left(d\theta^2+\sin^2\theta ~d\phi^2\right).
   \label{metr1}
\end{eqnarray}
The global static coordinates (\ref{coord1}) cover the whole anti-de~Sitter
space (except the coordinate singularities at $\chi=0$ and $\theta=0,\pi$), 
the time coordinate $t$ has a period of $2\pi\alpha$.
2) {\em Local static coordinates}. Define
\begin{eqnarray}
   \left\{\begin{array}{l}
   V=\left(r^2-\alpha^2\right)^{1/2}\sinh{t\over\alpha}\\
   W=r\cosh\theta\\
   X=\left(r^2-\alpha^2\right)^{1/2}\cosh{t\over\alpha}\\ 
   Y=r\sinh\theta\cos\phi\\
   Z=r\sinh\theta\sin\phi
   \end{array}
   \right.,
   \label{coord2}
\end{eqnarray}
where $-\infty<t<\infty$, $r>\alpha$, $0<\theta<\infty$, and
$0\leq\phi<2\pi$. Then the anti-de~Sitter metric can be written 
(in a Schwarzschild like form) as
\begin{eqnarray}
   ds^2=-\left({r^2\over\alpha^2}-1\right) dt^2+
   \left({r^2\over\alpha^2}-1\right)^{-1} dr^2+
   r^2\left(d\theta^2+\sinh^2\theta d\phi^2\right).
   \label{metr2}
\end{eqnarray}
The local static coordinates (\ref{coord2}) cover only the region with
$|V|<X$ and $W>0$ in anti-de~Sitter space. 
3) {\em Non-stationary coordinates}. Define
\begin{eqnarray}
   \left\{\begin{array}{l}
   V=\alpha\cos t\cosh\chi\\
   W=\alpha\sin t\\
   X=\alpha\cos t\sinh\chi\cos{\theta}\\
   Y=\alpha\cos t\sinh\chi\sin{\theta}\cos\phi\\
   Z=\alpha\cos t\sinh\chi\sin{\theta}\sin\phi
   \end{array}\right.,
   \label{coord3}
\end{eqnarray}
where $-\pi/2<t<\pi/2$, $0<\chi<\infty$, 
$0<\theta<\pi$, and $0\leq\phi<2\pi$.
Then the anti-de~Sitter metric can be written (in an open cosmological form) as
\begin{eqnarray}
   ds^2=\alpha^2\left\{-dt^2+\cos^2 t\left[d\chi^2+\sinh^2\chi
   \left(d\theta^2+\sin^2\theta ~d\phi^2\right)\right]\right\}.
   \label{metr3}
\end{eqnarray}
The non-stationary coordinates (\ref{coord3}) cover the region with
$V>0$ and $|W|<\alpha$, but they can by extended to the region with
$V<0$ and $|W|<\alpha$ by the transformations $t\rightarrow \pi+t$,
$\chi\rightarrow-\chi$, $\theta\rightarrow \pi-\theta$, and
$\phi\rightarrow\pi+\phi$. 
4) The non-stationary coordinates
can also be extended to the regions with $|W|>\alpha$. Define
\begin{eqnarray}
   \left\{\begin{array}{l}
   V=\alpha\sinh t\sinh\chi\\
   W=-\alpha\cosh\chi\\
   X=\alpha\cosh t\sinh\chi\cos\theta\\
   Y=\alpha\cosh t\sinh\chi\sin\theta\cos\phi\\
   Z=\alpha\cosh t\sinh\chi\sin{\theta}\sin\phi
   \end{array}\right.,
   \label{coord3a}
\end{eqnarray}
where $-\infty<t<\infty$, $0<\chi<\infty$, $0<\theta<\pi$, and $0\leq\phi
<2\pi$. The coordinates $(t,\chi,\theta,\phi)$ cover the region with
$W<-\alpha$, where the anti-de~Sitter metric can be written as
\begin{eqnarray}
   ds^2=\alpha^2\left[-\sinh^2\chi ~dt^2+d\chi^2+\cosh^2 t
   \sinh^2\chi\left(d\theta^2+\sin^2\theta~d\phi^2\right)\right].
   \label{metr3a}
\end{eqnarray}

Anti-de~Sitter space is multiply connected and has CTCs everywhere. For
example, in the global static coordinates (\ref{coord1}), 
the world line with $\chi=
{\rm const}$, $\theta={\rm const}$, and $\phi={\rm const}$ (which is the
intersection of the hyperbola given by Eq.~(\ref{ades}) with the
surface with $X={\rm const}$, $Y={\rm const}$, and $Z={\rm const}$) is a CTC
with the proper period $2\pi\alpha\cosh\chi$. If we unfold the anti-de~Sitter
space along the time coordinate $t$ in the global static coordinates 
(then $t$ goes from $-\infty$ to $\infty$), we 
obtain the covering space of the anti-de~Sitter space, which is simply
connected with the topology
$R^4$ and does not contain CTCs anymore. (However, there is no Cauchy surface 
in this covering space \cite{haw73}.) The Penrose diagram of the covering 
space of anti-de~Sitter space is shown in Fig.~\ref{fig1}. 
Anti-de~Sitter space has maximum 
symmetry, which has one time translation Killing vector, three space 
rotation Killing vectors, and six boost Killing 
vectors. 
In the local static coordinates in (\ref{coord2}), $\partial/\partial t$ is
a boost Killing vector. By the continuation
\begin{eqnarray}
   t\rightarrow l-i{\pi\over2}\alpha, \hskip 1.0cm r\rightarrow \tilde{t},
   \label{cont}
\end{eqnarray}
where $-\infty< l<\infty$ and $-\alpha<\tilde{t}<\alpha$, the local static
coordinates can be extended to the region with $V>|X|$ on the 
hyperbola defined by Eq.~(\ref{ades}), where the boost Killing vector becomes
$\partial/\partial l$ and the anti-de~Sitter metric can be written as
\begin{eqnarray}
   ds^2=-\left(1-{\tilde{t}^2\over\alpha^2}\right)^{-1} d\tilde{t}^2+
   \left(1-{\tilde{t}^2\over\alpha^2}\right) dl^2+
   \tilde{t}^2\left(d\theta^2+\sin^2\theta d\phi^2\right).
   \label{metr4}
\end{eqnarray}
In the covering space of the anti-de~Sitter space, if we identify all
points related by boost transformations, then we obtain a {\em Misner-like
anti-de~Sitter space}. With this identification, 
there are CTCs in the region with $|V|<|X|$ but no CTCs in the region
with $|V|>|X|$. On the boundary $|V|=|X|$, there are closed null curves. 
$|V|=|X|$ is the chronology horizon. (See Fig.~\ref{fig2} for the causal
structure of Misner-like anti-de~Sitter space.)

A coordinate system $(t,x,y,z)$ covering the 
region $V+X>0$ of anti-de~Sitter space can
also be found, which are given by
\begin{eqnarray}
   \left\{\begin{array}{l}
   V=\alpha\cosh\psi+{1\over2\alpha}e^{-\psi}\left(y^2+z^2-t^2\right)\\
   W=t\\
   X=\alpha\sinh\psi-{1\over2\alpha}e^{-\psi}\left(y^2+z^2-t^2\right)\\
   Y=y\\
   Z=z
   \end{array}\right.,
   \label{coord5}
\end{eqnarray}
where $-\infty<t,\psi,y,z<\infty$. With these coordinates, the anti-de~Sitter
metric can be written as 
\begin{eqnarray}
   ds^2=-(dt-td\psi)^2+\alpha^2 d\psi^2+(dy-yd\psi)^2+(dz-zd\psi)^2.
   \label{metr5}
\end{eqnarray}
$\partial/\partial\psi$ is a boost Killing vector. In the covering space
of anti-de~Sitter space, if the points 
$(t,\psi,y,z)$ are identified with $(t,\psi+2n\pi,y,z)$ ($n=\pm 1,\pm 2,...$),
we obtain a Misner-like anti-de~Sitter space. In this space there are CTCs
in the regions with $t^2>\alpha^2+y^2+z^2$, but no CTCs in the region with
$t^2<\alpha^2+y^2+z^2$.

In 1993, Li, Xu, and Liu \cite{li93}
constructed a complex space $S^1\times R^3$ with
the metric 
\begin{eqnarray}
   ds^2=(dw-wd\psi)^2+d\psi^2+(y-yd\psi)^2+(z-zd\psi)^2,
   \label{metr6}
\end{eqnarray}
where $\psi$, $y$, and $z$ are real but $w$ is complex, and $\psi$ has a 
period $2\pi$ [i.e. $(w,\psi,y,z)$ are identified with $(w,\psi+2n\pi,y,z)$
where $n=\pm 1\pm 2,...$]. They showed that this space is a solution of
the vacuum complex Einstein equation with a negative cosmological constant 
$\Lambda=-3$. Here we find that, the Lorentzian section of the metric in
(\ref{metr6}) (i.e. let
$w=it$) is just the anti-de~Sitter metric in (\ref{metr5}) with
$\alpha=1$ (i.e. $\Lambda=-3$). Thus
the Lorentzian section of the complex space of Li, Xu, and Liu is just
the Misner-like anti-de~Sitter space obtained from the covering
space of the anti-de~Sitter space by identifying points related by 
boost transformations.

\section{Self-consistent Vacuum in Misner-like Anti-de~Sitter Space}
Usually there is no well-defined quantum field theory in a spacetime
with CTCs. However the problem can be worked using Hawking's Euclidean
quantization procedure \cite{haw79,haw95}. Alternatively, in the case where a
covering space exists, we can
do it in the covering space with the method of images. In fact, in
most cases where the renormalized energy-momentum tensor
in spacetimes with CTCs has been calculated,
this method has been used (for the
theoretical basis for the method of images see Ref. \cite{fro91} and
references cited therein). Thus we will begin by using 
this method to deal with quantum field theory
in anti-de~Sitter space and Misner-like anti-de~Sitter space.

Anti-de~Sitter space is conformally flat. With the transformation
$\chi^\prime=2\arctan e^\chi-{1\over2}\pi$, the anti-de~Sitter metric in
Eq.~(\ref{metr1}) can be written as
\begin{eqnarray}
   ds^2=\alpha^2\cosh^2\chi ~d\tilde{s}^2=
   \alpha^2\cosh^2\chi \left[-dt^2+d{\chi^{\prime}}^2+\sin^2
   \chi^{\prime}\left(d\theta^2+\sin^2\theta ~d\phi^2\right)\right],
\end{eqnarray}
where $d\tilde{s}^2=-dt^2+d{\chi^{\prime}}^2+\sin^2
\chi^{\prime}\left(d\theta^2+\sin^2\theta ~d\phi^2\right)$ is the metric of the
Einstein static universe. With more transformation $r=\sin\chi^\prime/\left[
2\left(\cos t+\cos\chi^\prime\right)\right]$ and $t^\prime=\sin t/\left[
\left(\cos t+\cos\chi^\prime\right)\right]$, the metric of anti-de~Sitter 
space can be written as
\begin{eqnarray}
   ds^2=\Omega^2d\overline{s}^2=
   \Omega^2\left[-{dt^{\prime}}^2+dr^2+r^2
   \left(d\theta^2+\sin^2\theta ~d\phi^2\right)\right],
   \label{conf1}
\end{eqnarray}
where $d\overline{s}^2=-{dt^{\prime}}^2+dr^2+r^2
\left(d\theta^2+\sin^2\theta ~d\phi^2\right)$ is just the Minkowski metric, 
and $\Omega^2$ is given by
\begin{eqnarray}
   \Omega^2=4\alpha^2\cosh^2\chi\left(\cos t+\cosh\chi\right)^2.
\end{eqnarray}
Eq.~(\ref{conf1}) demonstrates that anti-de~Sitter space is conformally 
flat.

For a massless conformally coupled scalar field in a
conformally flat spacetime, there exists a conformal vacuum whose Hadamard
function $G^{(1)}(X,X^\prime)$ is related to the Hadamard function
$\overline{G}^{(1)}(X,X^\prime)$ of the corresponding vacuum of the massless
conformally coupled scalar field in flat Minkowski space via \cite{bir82}
\begin{eqnarray}
   G^{(1)}(X,X^\prime)=\Omega^{-1}(X)\overline{G}^{(1)}
   (X,X^\prime)\Omega^{-1}(X^\prime).
   \label{had}
\end{eqnarray}
The corresponding renormalized stress-energy tensors are related via
\begin{eqnarray}
   \langle T_a^{~b}\rangle_{\rm ren}=\Omega^{-4}\langle\overline
   {T}_a^{~b}\rangle_{\rm ren}+{1\over16\pi^2}\left[{1\over9}
   a_1 ~^{(1)}H_a^{~b}+2a_3 ~^{(3)}H_a^{~b}\right],
   \label{ener}
\end{eqnarray}
where
\begin{eqnarray}
   &&^{(1)}H_{ab}=2\nabla_a\nabla_bR-2g_{ab}\nabla^c\nabla_cR
   -{1\over2}R^2g_{ab}+2RR_{ab}, \\
   &&^{(3)}H_{ab}=R_a^{~c}R_{cb}-{2\over3}RR_{ab}-{1\over2}R_{cd}R^{cd}
   g_{ab}+{1\over4}R^2g_{ab},
\end{eqnarray}
and for scalar field we have $a_1={1\over120}$ and $a_3=-{1\over360}$
\cite{bir82}. [The sign before $1/16\pi^2$ is positive here because we
are using signature $(-,+,+,+)$]. For
anti-de~Sitter space we have $R_{ab}=\Lambda g_{ab}$, $R=4\Lambda$,
and thus $^{(1)}H_{ab}=0$, $^{(3)}H_{ab}={1\over3}\Lambda^2g_{ab}
={3\over \alpha^4}g_{ab}$.
Inserting these into Eq.~(\ref{ener}), we have
\begin{eqnarray}
   \langle T_a^{~b}\rangle_{\rm ren}=\Omega^{-4}\langle\overline
   {T}_a^{~b}\rangle_{\rm ren}-{1\over960\pi^2 \alpha^4}\delta_a^{~b}.
   \label{ener1}
\end{eqnarray}

In Minkowski spacetime, for the massless conformally coupled scalar 
field with the Minkowski vacuum, the Hadamard function is
\begin{eqnarray}
   \overline{G}_{\rm M}^{(1)}(X,X^\prime)={1\over 2\pi^2}{1\over
   -\left(t^{\prime}-{t^\prime}^\prime\right)^2+r^2+{r^\prime}^2
   -2 r r^\prime\cos\Theta_2},
   \label{mg}
\end{eqnarray}
where $\cos\Theta_2=\cos\theta\cos\theta^\prime+
\sin\theta\sin\theta^\prime\cos(\phi-\phi^\prime)$.
The corresponding renormalized stress-energy tensor of the Minkowski
vacuum is
\begin{eqnarray}
   \langle\overline{T}_{ab}\rangle_{\rm ren}=0.
   \label{me}
\end{eqnarray}
Inserting Eq.~(\ref{mg}) into Eq.~(\ref{had}), we get the Hadamard function
for the massless conformally coupled scalar field in the conformal
Minkowski vacuum in anti-de~Sitter space
\begin{eqnarray}
   G_{\rm CM}^{(1)}(X,X^\prime)={1\over4\pi^2\alpha^2}{1\over\cos
   \left[(t-t^\prime)/\alpha\right]
   \cosh \chi\cosh \chi^\prime-1-\sinh \chi\sinh \chi^\prime
   \cos\Theta_2},
   \label{adg}
\end{eqnarray}
where $(t,\chi,\theta,\phi)$ are the global static coordinates of
anti-de~Sitter space. Clearly $G_{\rm CM}^{(1)}$ satisfies the periodic
boundary condition
\begin{eqnarray}
   G_{\rm CM}^{(1)}(t,\chi,\theta,\phi;t^\prime,\chi^\prime,
   \theta^\prime,\phi^\prime)=G_{\rm CM}^{(1)}
   (t+2n\pi\alpha,\chi,\theta,\phi;t^\prime,\chi^\prime,
   \theta^\prime,\phi^\prime),
   \label{boun}
\end{eqnarray}
thus it is a suitable Hadamard function in anti-de~Sitter
space which has CTCs everywhere. Inserting Eq.~(\ref{me}) into
Eq.~(\ref{ener1}), we get the renormalized stress-energy tensor for
the massless and conformally coupled scalar field in the conformal Minkowski
vacuum in anti-de~Sitter space
\begin{eqnarray}
   \langle T_{ab}\rangle_{\rm ren}=-{1\over 960\pi^2\alpha^4}g_{ab},
   \label{ener2}
\end{eqnarray}
which is the same as that for de~Sitter space with radius $\alpha$.

If we insert the energy-momentum tensor in Eq.~(\ref{ener2}) into the
semiclassical Einstein equations 
\begin{eqnarray}
   G_{ab}+\Lambda g_{ab}=8\pi\langle T_{ab}\rangle_{\rm ren},
   \label{ein}
\end{eqnarray}
and recall that for anti-de~Sitter space we have $G_{ab}=R_{ab}-{1\over2}R
g_{ab}={3\over \alpha^2}g_{ab}$, we find that the semiclassical 
Einstein equations are satisfied if and only if
\begin{eqnarray}
   \Lambda+{3\over\alpha^2}+{1\over120\pi \alpha^4}=0.
   \label{eina}
\end{eqnarray}
If $\Lambda=0$, the two solutions to Eq.~(\ref{eina}) are
$\alpha=\infty$, which corresponds to Minkowski space; and $\alpha=
i(360\pi)^{-1/2}$, which corresponds to a de~Sitter space
with radius $|\alpha|=(360\pi)^{-1/2}$.
Thus, if the bare cosmological constant $\Lambda=0$, there is no 
self-consistent anti-de~Sitter space (though there are a
self-consistent de~Sitter space with radius $|\alpha|
=(360\pi)^{-1/2}$ \cite{got82,got98} and a self-consistent Minkowski
space). If $\Lambda<0$, the two solutions to Eq.~(\ref{eina}) are 
\begin{eqnarray}
   \alpha^2=-{3\over2\Lambda}\left(1+\sqrt{1-{\Lambda\over270\pi}}\right)>0,
   \label{sol1}
\end{eqnarray}
which corresponds to an anti-de~Sitter space with radius $\alpha$, and
\begin{eqnarray}
   \alpha^2=-{3\over2\Lambda}\left(1-\sqrt{1-{\Lambda\over270\pi}}\right)<0,
   \label{sol2}
\end{eqnarray}
which corresponds to a de~Sitter space with radius $|\alpha|$. 
(If $\Lambda>0$, the two solutions both correspond to de~Sitter spaces
\cite{got98}.) 
It is interesting to note that if $\Lambda<0$ there are two self-consistent
spaces, one of which is an anti-de~Sitter space, and the other is a 
de~Sitter space.  
Eq.~(\ref{eina}) tells us that, if $\Lambda<0$ and 
$\alpha^2<0$, we have $|\alpha|<(360\pi)^{-1/2}$, 
which implies that the self-consistent de~Sitter space
supported by a negative cosmological constant has a sub-Planckian radius. 
(See Sec.~V for further discussion.)  

For the Misner-like anti-de~Sitter space which is obtained from the
covering space of anti-de~Sitter space by identifying points related by
boost transformations, as in the case of Misner-like de~Sitter space
\cite{got98}, it is easily to show that the adapted 
conformal Minkowski vacuum is
not a self-consistent quantum state for the massless conformally coupled
scalar field. The renormalized stress-energy tensor of the adapted conformal
Minkowski vacuum diverges as the chronology horizon is approached. But, as
in the cases of Misner space \cite{li98} and Misner-like de~Sitter space
\cite{got98}, we can show that an adapted conformal Rindler vacuum is
a self-consistent vacuum state for a massless conformally coupled 
scalar field in Misner-like anti-de~Sitter space. To do so, it will be
more convenient to write the anti-de~Sitter metric in the local static
coordinates [Eqs.~(\ref{coord2}) and (\ref{metr2})]
and the Minkowski metric in Rindler coordinates
\begin{eqnarray}
   d\overline{s}^2=-\xi^2d\eta^2+d\xi^2+dy^2+dz^2,
   \label{rin1}
\end{eqnarray}
where the Rindler coordinates $(\eta,\xi,y,z)$ are defined by
\begin{eqnarray}
   \left\{\begin{array}{l}
   t=\xi\sinh\eta\\ 
   x=\xi\cosh\eta\\
   y=y\\
   z=z
   \end{array}\right.,
   \label{rin2}
\end{eqnarray}
where $(t,x,y,z)$ are the Cartesian coordinates in Minkowski spacetime.
With the transformation
\begin{eqnarray}
   \left\{\begin{array}{l}
   \eta={t\over\alpha}\\
   \xi={\sqrt{r^2/\alpha^2-1}\over (r/\alpha)\cosh\theta-1}\\
   y={(r/\alpha)\sinh\theta\cos\phi\over (r/\alpha)\cosh\theta-1}\\
   z={(r/\alpha)\sinh\theta\sin\phi\over (r/\alpha)\cosh\theta-1}
   \end{array}\right.,
   \label{rin3}
\end{eqnarray}
where $(t,r,\theta,\phi)$ are the local static coordinates in 
Eq.~(\ref{coord2}), the anti-de~Sitter metric (\ref{metr2}) can be written
as
\begin{eqnarray}
   ds^2=\Omega^2d\overline{s}^2=\Omega^2\left(-\xi^2 d\eta^2+
   d\xi^2+dy^2+dz^2\right),
   \label{adsr}
\end{eqnarray}
here $\Omega^2$ is
\begin{eqnarray}
   \Omega^2=\alpha^2\left[(r/\alpha)\cosh\theta-1\right]^2.
\end{eqnarray}
With the conformal relation between anti-de~Sitter space and Rindler space
given by Eqs.~(\ref{rin3}) and (\ref{adsr}), the time coordinate $t$ of
the anti-de~Sitter space in local static coordinates is mapped to the 
Rindler time coordinate $\eta$. Construct a Misner space with all 
$(\eta+n\eta_0,\xi,y,z)$ ($n=0,\pm1,...$) identified, then the map 
given by Eq.~(\ref{rin3}) gives rise to a Misner-like anti-de~Sitter
space with all
$(t+nt_0,r,\theta,\phi)$ ($n=0,\pm1,...$) identified, where
\begin{eqnarray}
   t_0=\alpha\eta_0.
   \label{t0}
\end{eqnarray}
Eqs.~(\ref{rin3})-(\ref{t0}) give a natural conformal map
between Misner space and Misner-like anti-de~Sitter space.

For a massless conformally coupled scalar field in Misner space, the
Hadamard function for the adapted Rindler vacuum is \cite{li98}
\begin{eqnarray}
   G^{(1)}(X,X^\prime)={1\over2\pi^2}\sum_{n=-\infty}^{\infty}
   {\gamma\over\xi\xi^\prime
   \sinh\gamma\left[-(\eta-\eta^\prime+n\eta_0)^2+\gamma^2\right]},
   \label{ring}
\end{eqnarray}
where $\gamma$ is defined by
\begin{eqnarray}
   \cosh\gamma={\xi^2+{\xi^\prime}^2+(y-y^\prime)^2+(z-z^\prime)^2
   \over 2\xi\xi^\prime}.
\end{eqnarray}
The corresponding renormalized stress-energy tensor of the adapted Rindler
vacuum is \cite{li98}
\begin{eqnarray}
   \langle T_\mu^{~\nu}\rangle_{\rm R,ren}={1\over1440\pi^2\xi^4}
   \left[\left({2\pi\over \eta_0}\right)^4-1\right]
   \left(\begin{array}{cccc}
   -3&0&0&0\\
   0&1&0&0\\
   0&0&1&0\\
   0&0&0&1
   \end{array}
   \right),
   \label{E69}
\end{eqnarray}
which is expressed in the Rindler coordinates. We see that $\langle T_{\mu}^
{~\nu}\rangle_{\rm R,ren}$ (also $\langle T_{\mu\nu}
\rangle_{\rm R,ren}\langle T^{\mu\nu}
\rangle_{\rm R,ren}$) diverges at the chronology horizon where 
$\xi=0$ unless $\eta_0=2\pi$. If $\eta_0=2\pi$, however, we have $\langle T_{\mu}^
{~\nu}\rangle_{\rm R,ren}=0$ throughout the Misner space (though Rindler
coordinates cover only a quadrant in Misner space, the results can be
analytically extended to the whole Misner space \cite{li98,got98} where $\langle
T_\mu^{~\nu}\rangle_{\rm R,ren}$ is also zero, see \cite{got98} for further
discussion).

Inserting Eq.~(\ref{ring}) into Eq.~(\ref{had}), we obtain the Hadamard
function for the adapted conformal Rindler vacuum of the massless 
and conformally
coupled scalar field in Misner-like anti-de~Sitter space
\begin{eqnarray}
   G_{\rm CR}^{(1)}(X,X^\prime)={1\over2\pi^2}\sum_{n=-\infty}^{\infty}
   {\gamma\over\sinh\gamma\sqrt{(r^2/\alpha^2-1)({r^\prime}^2/\alpha^2-1)}
   ~\left[-(t-t^\prime+nt_0)^2+ \alpha^2\gamma^2\right]},
   \label{E79}
\end{eqnarray}
where $\gamma$ is written in $(t,r,\theta,\phi)$ as
\begin{eqnarray}
   \cosh\gamma&=&{1\over\sqrt{(r^2/\alpha^2-1)({r^\prime}^2/\alpha^2-1
   )}} \nonumber\\
   &&\times\left\{
   1-{rr^\prime\over \alpha^2}[\cosh\theta\cosh\theta^\prime-\sinh\theta
   \sinh\theta^\prime\cos(\phi-\phi^\prime)]\right\}.
   \label{E80}
\end{eqnarray}
Clearly the Hadamard function in Eq.~(\ref{E79}) satisfies the periodic
boundary condition
\begin{eqnarray}
   G^{(1)}(t,r,\theta,\phi)=G^{(1)}(t+nt_0,r,\theta,\phi),
\end{eqnarray}
where $n=0,\pm1,...$, thus it defines a reasonable quantum state in the
Misner-like anti-de~Sitter space. Inserting Eq.~(\ref{E69}) into
Eq.~(\ref{ener1}), we obtain the renormalized stress-energy tensor of
the adapted conformal Rindler vacuum of the conformally coupled scalar
field in Misner-like anti-de~Sitter space
\begin{eqnarray}  
   \langle T_\mu^{~\nu}\rangle_{\rm CR,ren}={1\over1440\pi^2
   \alpha^4\left(r^2/\alpha^2-1\right)^2}\left[\left({2\pi\alpha
   \over t_0}\right)^4-1\right]
   \left(\begin{array}{cccc}
   -3&0&0&0\\
   0&1&0&0\\
   0&0&1&0\\
   0&0&0&1
   \end{array}
   \right)-{1\over960\pi^2 \alpha^4}\delta_{\mu}^{~\nu},
   \label{E73}
\end{eqnarray}
where the coordinate system is the local static coordinate system 
$(t,r,\theta,\phi)$. Again, $\langle T_\mu^{~\nu}\rangle_{\rm CR,ren}$
(also $\langle T_{\mu\nu}\rangle_{\rm CR,ren} 
\langle T^{\mu\nu}\rangle_{\rm CR,ren}$)
diverges at the chronology horizon $r=\alpha$, unless $t_0=2\pi\alpha$.
But, if
\begin{eqnarray}
   t_0=2\pi\alpha,
   \label{con2}
\end{eqnarray}
we get a renormalized stress-energy tensor which is regular throughout
the Misner-like anti-de~Sitter space
\begin{eqnarray}
   \langle T_{ab}\rangle_{\rm CR,ren}=-{1\over960\pi^2 \alpha^4}g_{ab},
   \label{E75}
\end{eqnarray}
which is exactly the same as that of de~Sitter space and anti-de~Sitter
space. [Though the local coordinates $(t,r,\theta,\phi)$ cover only 
the region with $|V|<X$ in anti-de~Sitter space, 
the results can be easily extended
to the whole Misner-like anti-de~Sitter space (where 
$\langle T_{ab}\rangle_{\rm CR,ren}$ is finite and 
given by Eq.~(\ref{E75})), as in the cases of Misner 
space and Misner-like de~Sitter space \cite{li98,got98}.]

Similarly, the Misner-like anti-de~Sitter space solves the
semiclassical Einstein equation with a negative cosmological 
constant $\Lambda$ and the energy-momentum
tensor in Eq.~(\ref{E75}) (and thus is self-consistent) if
$\alpha^2=-{3\over2\Lambda}\left(1+\sqrt{1-{\Lambda\over 270\pi}}\right)$.

\section{Transition between De~Sitter and Anti-de~Sitter Spaces}
Phase transitions play important roles in the evolution of the early universe.
With phase transitions various bubbles could 
form; inside and outside the bubbles
the spacetimes have different stress-energy tensors and thus 
are described by different spacetime metrics \cite{col77,col80}. 
The inside and outside of a bubble are separated by a
wall --- a spacetime structure which can be approximately treated as a
three-dimensional hypersurface. 
Usually it is assumed that the outside of the bubble
is in a state dominated by a positive cosmological constant 
at GUT (or Planckian) scale. Thus the spacetime
metric outside the bubble
is well approximated with a de~Sitter metric. Inside the bubble, the 
cosmological constant
could be zero and the stress-energy tensor could be zero
thus the spacetime inside would be Minkowskian --- which is a 
model of inflation decaying through a first order
phase transition in the old inflation theory \cite{gut81}; or, inside
the bubble the cosmological constant
could also be positive and at GUT (or Planckian) scale
so the spacetime inside the bubble
is still inflating, but after a while the cosmological constant 
falls off a plateau and evolves classically 
to zero and the universe inside the bubble enters 
a hot Big Bang
phase --- which is a model of transition from inflation to an open Big Bang 
cosmology through
a second order phase transition \cite{got82,got86,lin82,alb82}. But, 
either via the first order phase transition or the second order
phase transition, as another
alternative, the inside of the bubble could become dominated by a negative
cosmological constant instead and thus the spacetime inside the bubble would be 
described with an anti-de~Sitter metric. In this paper we are interested in this
case since anti-de~Sitter space has CTCs. 

In this paper we will discuss bubbles that are pre-existing rather than ones
that form by quantum tunneling. We thus only consider how to glue
a de~Sitter space and an anti-de~Sitter space together at a boundary
(i.e. at a wall), and we investigate
the causal structure of the spacetime so obtained. 

The conditions for two spacetimes to be glued together along a wall 
(so that the Einstein
equations are satisfied at the wall) are \cite{mis73}: 1) the metrics 
on the wall induced
from the spacetimes at the two sides agree; 2) the surface 
stress-energy tensor of the wall
defined by $S_\alpha^{~\beta}=\lim_{\epsilon\rightarrow 0}\int_{-\epsilon}^
\epsilon T_{\alpha}^{~\beta}dn$ should satisfy (in the Gaussian normal
coordinates near the wall)
\begin{eqnarray}
   S_n^{~n}=S_n^{~i}=0,\hskip 1.0cm
   S_i^{~j}=-{n^an_a\over8\pi}\left(\gamma_i^{~j}-
   \gamma\delta_i^{~j}\right),
   \label{join}
\end{eqnarray}
where $n^a$ is the normal vector of the hypersurface $\Sigma$ of the wall
($n^an_a=1$ if the $\Sigma$ is timelike; $n^an_a=-1$ if the wall is spacelike.
Regarding $\Sigma$ as a hypersurface embedded in either the spacetime inside 
it, or the spacetime outside it, by the first condition, $\Sigma$ should
be either timelike or spacelike in both. Here we do 
not consider the case of a null wall),
and
\begin{eqnarray}
   \gamma_{ab}\equiv[K_{ab}]\equiv K_{ab}^+ -K_{ab}^-
   \label{join1}
\end{eqnarray}
is the difference of the
extrinsic curvature of $\Sigma$ embedded in the spacetimes at the two sides
of the wall \cite{mis73}, and $\gamma\equiv \gamma_a^{~a}$. 
The definition for the extrinsic curvature 
$K_{ab}$
is $K_{ab}=\nabla_a n_b$ 
where we have treated $n^a$ as an vector field in
the neighborhood of $\Sigma$ extended from the normal vector defined only
on $\Sigma$. In the Gaussian normal coordinates of $\Sigma$, the components
of $K_{ab}$ can be written as
\begin{eqnarray}
   K_{\mu\nu}={1\over2}{\partial h_{\mu\nu}\over \partial n}.
\end{eqnarray}
(Here we use the definition of the extrinsic 
curvature $K_{ab}$ with an opposite sign as that used in \cite{mis73}.)
The evolution of the wall is governed by
\begin{eqnarray}
   S^{ij}_{~~|j}+[T^{in}]=0,
   \label{evol}
\end{eqnarray}
where ``$|j$'' denotes the covariant derivative associated with the 
three-metric $h_{ij}$ on $\Sigma$. (Here the Greek letters $\mu,\nu,...$
label vectors and tensors in the four-dimensional spacetime, the Latin
letters $i,j,...$ label vectors and tensors in the three-dimensional
space $\Sigma$.)

De~Sitter space is a hyperbola
\begin{eqnarray}
   -V^2+W^2+X^2+Y^2+Z^2=\beta^2,
   \label{des}
\end{eqnarray}
embedded in a five-dimensional space $R^5$ with the metric
\begin{eqnarray}
   ds^2=-dV^2+dW^2+dX^2+dY^2+dZ^2.
   \label{des1}
\end{eqnarray}
De~Sitter space is a solution of the vacuum Einstein equations with a
positive cosmological constant $\Lambda=3/\beta^2$, 
which has maximal symmetry (it has six space rotation Killing vectors
and four boost Killing vectors).
There are various coordinate systems for de~Sitter space 
\cite{sch56,haw73,bir82}. For our purpose here, the following two
coordinate systems are convenient:       
1). Define
\begin{eqnarray}
   \left\{\begin{array}{l}
   V=\beta\sinh t\cosh\chi\\
   W=\beta\cosh t\\
   X=\beta\sinh t\sinh\chi\cos\theta\\
   Y=\beta\sinh t\sinh\chi\sin\theta\cos\phi\\
   Z=\beta\sinh t\sinh\chi\sin\theta\sin\phi
   \end{array}\right.,
   \label{dco2}
\end{eqnarray}
where $0<t<\infty$, $0<\chi<\infty$, $0<\theta<\pi$, and $0\leq\phi<2\pi$.
The coordinates $(t,\chi,\theta,\phi)$ in (\ref{dco2}) cover the region
on the hyperbola (\ref{des})
with $V>0$ and $W>\beta$, where the de~Sitter metric can be written as
\begin{eqnarray}
   ds^2=\beta^2\left\{-dt^2+\sinh^2 t\left[d\chi^2+
   \sinh^2\chi\left(d\theta^2+\sin^2\theta~d\phi^2\right)\right]\right\}.
   \label{dmetr2}
\end{eqnarray}
The section $t={\rm constant}$ in de~Sitter space is an open (i.e. negatively
curved), homogeneous, and isotropic space. The metric in (\ref{dmetr2})
could describe an open inflation \cite{got82,got86}.
The coordinates $(t,\chi,\theta,\phi)$ in (\ref{dco2}) can be extended to
the region with $V<0$ and $W>\beta$ by the reflection $t\rightarrow-t$
and $\chi\rightarrow-\chi$, $\theta\rightarrow\pi-\theta$, and 
$\phi\rightarrow\pi+\phi$. 2) The coordinates in (\ref{dco2}) could be extended 
to the region with $|W|<\beta$. Define
\begin{eqnarray}
   \left\{\begin{array}{l}
   V=\beta\sinh t\cos\chi\\
   W=\beta\sin\chi\\
   X=\beta\cosh t\cos\chi\cos\theta\\
   Y=\beta\cosh t\cos\chi\sin\theta\cos\phi\\
   Z=\beta\cosh t\cos\chi\sin\theta\sin\phi
   \end{array}\right.,
   \label{dco1}
\end{eqnarray}
where $-\infty<t<\infty$, $-\pi/2<\chi<\pi/2$, $0<\theta<\pi$,
and $0\leq\phi<2\pi$,
then the de~Sitter metric can be written as
\begin{eqnarray}
   ds^2=\beta^2\left[-\cos^2\chi~dt^2+ d\chi^2+
   \cosh^2 t\cos^2\chi~\left(d\theta^2+\sin^2\theta~d\phi^2
   \right)\right].
   \label{dmetr1}
\end{eqnarray}
The coordinates $(t,\chi,\theta,\phi)$ cover the region with $|W|<\beta$
in the de~Sitter space as a hyperbola given by Eq.~(\ref{des}).   
The Penrose diagram of de~Sitter space is
shown in Fig.~\ref{fig3}.

By gluing anti-de~Sitter space onto de~Sitter space we can obtain various
spacetimes with CTCs. These
spacetimes with CTCs differ from the anti-de~Sitter space with
CTCs by the fact that in these glued spacetimes there are regions without
CTCs which are separated from the regions with CTCs by chronology
horizons, while in the usual anti-de~Sitter space CTCs exist everywhere.
Here we only show one typical example, which is
obtained by gluing an anti-de~Sitter space onto a de~Sitter
along a timelike hypersurface ($n^a n_a=1$). 
This spacetime could describe a bubble
of anti-de~Sitter space existing for all time in an eternal de~Sitter space.
 
Consider a de~Sitter space as a hyperbola described by Eq.~(\ref{des})
in the embedding space with the metric in Eq.~(\ref{des1}).
Cut this de~Sitter space along the hypersurface 
$\Sigma_1$ with $W=w_1>0$ and throw
away the part with $W>w_1$. Denote the part of the 
de~Sitter space with $W<w_1$ as $dS^-$. Then we have a de~Sitter 
space with a boundary $\Sigma_1$ at $W=w_1$. Suppose $w_1<\beta$,
then $\Sigma_1$ is timelike. In such a case, the 
hypersurface $\Sigma_1$ is a three-dimensional timelike hyperbola
with $-V^2+X^2+Y^2+Z^2=\beta^2-w_1^2>0$.
With the coordinates in (\ref{dco1}), $\Sigma_1$ is
at $\chi=\arcsin(w_1/\beta)$. The metric $h_{ab}$ on
$\Sigma_1$ induced from the de~Sitter metric is
\begin{eqnarray}
   ds_1^2=\left(\beta^2-w_1^2\right)\left[-dt^2
   +\cosh^2 t\left(d\theta^2+
   \sin^2\theta~d\phi^2\right)\right].
   \label{hab1}
\end{eqnarray}
The normal vector of $\Sigma$ is 
$n^a=\beta^{-1}(\partial/\partial\chi)^a$ ($n^an_a=1$).
The extrinsic curvature $K_{ab}^-$ of $\Sigma_1$ is
\begin{eqnarray}
   K_{ab}^-=-{w_1\over \beta\sqrt{\beta^2-w_1^2}}h_{ab}.
   \label{kab1}
\end{eqnarray}

Consider an anti-de~Sitter space as the hyperbola given by
Eq.~(\ref{ades}) in the embedding space with the metric in Eq.~(\ref{metr0}).
Cut the anti-de~Sitter space along the hypersurface $\Sigma_2$ with
$W=-w_2<0$, throw away the part with $W<-w_2$. 
We denote the anti-de~Sitter space with $W>-w_2$ as $AdS^+$. Then we have
an anti-de~Sitter space with $W>-w_2$ and a boundary $\Sigma_2$ 
at $W=-w_2$. Suppose $w_2>\alpha$, then $\Sigma_2$ is
timelike. 
In such a case, $\Sigma_2$ is a three-dimensional 
timelike hyperbola with $-V^2+X^2+Y^2+Z^2=w_2^2-\alpha^2>0$. With the 
coordinates $(t,\chi,\theta,\phi)$ in (\ref{coord3a}), the hypersurface
$\Sigma_2$ is at $\chi={\rm arccosh}(w_2/\alpha)$. 
The metric $h_{ab}$ on $\Sigma_2$ induced from the anti-de~Sitter metric
is
\begin{eqnarray}
   ds_3^2=\left(w_2^2-\alpha^2\right)\left[-dt^2+\cosh^2 t
   \left(d\theta^2+\sin^2\theta~d\phi^2\right)\right].
   \label{hab3}
\end{eqnarray}
The normal vector of
$\Sigma_2$ is $n^a=-\alpha^{-1}(\partial/\partial\chi)^a$ 
($n^an_a=1$). 
The extrinsic curvature $K_{ab}^+$ of $\Sigma_2$ is
\begin{eqnarray}
   K_{ab}^+=-{w_2\over\alpha\sqrt{w_2^2-\alpha^2}}h_{ab}.
   \label{kab3}
\end{eqnarray}

To glue the anti-de~Sitter space and the de~Sitter space together, let us
identify $\Sigma_1$ with $\Sigma_2$ by identifying their
coordinates $(t,\theta,\phi)$. The spacetime so obtained is
schematically shown in Fig.~\ref{fig4}. In the sections with $V={\rm constant}$
in de~Sitter space and anti-de~Sitter space, an $S^3$ is glued together with an
$H^3$ at a cross-section $S^2$. From
Eqs.~(\ref{hab1}) and (\ref{hab3}), 
the metric $h_{ab}$ on $\Sigma_1=\Sigma_2\equiv\Sigma$ induced from de~Sitter
space and anti-de~Sitter space agree if and only if
\begin{eqnarray}
   w_1^2+w_2^2=\alpha^2+\beta^2.
   \label{ab}
\end{eqnarray}
By gluing two spacetimes along a hypersurface, usually a surface stress-energy
tensor is induced. The surface stress-energy tensor induced on the
hypersurface $\Sigma$  
is determined by the difference in the extrinsic curvature of $\Sigma_1$
which is embedded in the de~Sitter space and the extrinsic curvature
of $\Sigma_2$ which is embedded in 
the anti-de~Sitter space through
Eq.~(\ref{join}) with $n^an_a=1$. 
By inserting the extrinsic curvatures derived 
above into Eqs.~(\ref{join1}) and (\ref{join}), we obtain the surface
stress-energy tensor for $\Sigma$
\begin{eqnarray}
   S_{ab}=-{1\over 4\pi\mu}\left(\sqrt{1+{\mu^2\over\alpha^2}}-
   \sqrt{1-{\mu^2\over\beta^2}}\right)h_{ab},
   \label{sab1}
\end{eqnarray}
where $\mu=\sqrt{w_2^2-\alpha^2}=\sqrt{\beta^2-w_1^2}$, the metric $h_{ab}$
on the timelike $\Sigma$ is given by
\begin{eqnarray}
   ds^2=\mu^2\left[-dt^2+\cosh^2 t\left(d\theta^2+\sin^2\theta~d\phi^2
   \right)\right].
   \label{sab11}
\end{eqnarray}
The surface stress-energy tensor given by Eq.~(\ref{sab1}) is like a
positive three-dimensional cosmological constant. The timelike
hypersurface $\Sigma$ with the metric (\ref{sab11}) is a three-dimensional
de~Sitter space.
The Penrose diagram of the spacetime obtained by gluing $dS^-$
with $AdS^+$ along the timelike $\Sigma$ 
is shown in Fig.~\ref{fig5}. There are CTCs in the region of $dS^-$
with $W>-\beta$ and the whole $AdS^+$. But there are no CTCs in the
region of $dS^-$ with $W<-\beta$. The null hypersurface $W=-\beta$ is
the chronology horizon which separates the region with CTCs from that
without CTCs. (See Fig.~\ref{fig5}.)

Mathematically, a de~Sitter space can also 
be glued to an anti-de~Sitter space along a
spacelike hypersurface, the resultant spacetime has similar properties
as the example described above.
The surface stress-energy tensor induced on the spacelike hypersurface
is like a
negative cosmological constant in a three-dimensional Euclidean
space. The spacelike hypersurface 
is a three-dimensional hyperbola $H^3$. 
The causal structure of the spacetime obtained by gluing a de~Sitter
space to an anti-de~Sitter space along a spacelike hypersurface 
is the same as that of the spacetime obtained by gluing a de~Sitter
space to an anti-de~Sitter space along a timelike hypersurface.
There are CTCs in the region of $dS^-$
with $W>-\beta$ and the whole $AdS^+$. But there are no CTCs in the
region of $dS^-$ with $W<-\beta$. The null hypersurface $W=-\beta$ is
the chronology horizon which separates the region with CTCs from that
without CTCs.
For both cases, we have $T^{in}=0$ and $h^{ij}_{~~|j}=0$, thus the evolution 
equation (\ref{evol}) is satisfied automatically.

From Eqs.~(\ref{sab1}) we see that if $\mu\rightarrow0$,
i.e. if $\Sigma$ becomes null, we have $S_{ab}
\rightarrow 0$. Though the coordinates 
$(t,\chi,\theta,\phi)$ are singular at the null $\Sigma$ with $w_1=\beta$
and $w_2=\alpha$, we can show that as $\mu\rightarrow 0$, 
the scalars $S\equiv h^{ab}S_{ab}$ and $S^{ab}S_{ab}$
also vanish as $\mu\rightarrow 0$. (These conclusions also hold
if $\Sigma$ is spacelike.) However,
since the metric $h_{ab}$ on a null hypersurface is 
degenerate, the formalism discussed above cannot be used to the junction
at null hypersurfaces. Thus the case of null $\Sigma$ requires more
discussion. Here we do not discuss this complicated but interesting
issue. (For detail discussions on the junction conditions
at null hypersurfaces, see references \cite{cla87,bar91}).
 
\section{Self-consistent Vacua for Spacetimes with CTCs}
As discussed in Sec.~III, an anti-de~Sitter space with radius $\alpha$ is 
self-consistent (i.e. the semi-classical Einstein equations (\ref{ein})
are satisfied with
$\langle T_{ab}\rangle_{\rm ren}$ being the renormalized stress-energy
tensor of vacuum polarization) if the negative cosmological constant is
\begin{eqnarray}
   \Lambda_1=-{3\over\alpha^2}\left(1+{1\over360\pi\alpha^2}\right),
   \label{sola}
\end{eqnarray}
while a de~Sitter space with radius $\beta$ is self-consistent if the
cosmological constant is \cite{got98}
\begin{eqnarray}
   \Lambda_2={3\over\beta^2}\left(1-{1\over 360\pi\beta^2}\right).
   \label{solb}
\end{eqnarray}
$\Lambda_1$ is always negative. $\Lambda_2$ could be either positive
or negative, depending on the value of $\beta$. 
From Eq.~(\ref{solb}), if $\beta>(360\pi)^{-1/2}$, $\Lambda_2$ is positive;  
if $\beta<(360\pi)^{-1/2}$, $\Lambda_2$ is negative.
Thus, interestingly,
{\em a de~Sitter space with sub-Planckian radius could be self-consistent
only if the bare cosmological constant is negative}. 

From Eqs.~(\ref{sola}) and (\ref{solb}), we see that if 
$\beta<(360\pi)^{-1/2}$, $\Lambda_1=\Lambda_2$ if and only if
\begin{eqnarray}
    {1\over \beta^2}-{1\over\alpha^2}=360\pi.
    \label{solc}
\end{eqnarray}
This together with Eq.~(\ref{ab}) gives a self-consistent
spacetime which is obtained by gluing a de~Sitter space
to an anti-de~Sitter space with {\em a unique negative bare cosmological
constant throughout}. 
This could be realized since for a negative cosmological constant
there are two self-consistent solutions of the semi-classical Einstein
equations, one is anti-de~Sitter space, the other is de~Sitter space, and these two
could be glued together --- as
discussed in Sec.~III. 

The spacetime obtained by gluing a de~Sitter space with an 
anti-de~Sitter space as discussed in the last section, is a self-consistent
solution of the semi-classical Einstein equations if Eqs.~(\ref{sola})
and (\ref{solb}) are satisfied and on the wall separating the de~Sitter region
and the anti-de~Sitter region there is a surface stress-energy tensor given by
Eqs.~(\ref{sab1}). Since the Hadamard function is continuous
across the wall (the Hadamard function does not contain 
any derivatives of the metric), the
wall does not introduce any additional vacuum polarization effects. 

The above discussions of self-consistent solutions are for a massless
conformally coupled scalar field in de~Sitter/anti-de~Sitter spaces. The
results can be easily extended to other matter fields. If their vacua are
invariant under de~Sitter/anti-de~Sitter transformations, it could be 
expected that their renormalized stress-energy tensor should have the form
of ${\rm a ~constant}\times g_{ab}$. If there are many matter fields with
their vacua being invariant under de~Sitter/anti-de~Sitter transformations,
the renormalized stress-energy tensor could be written as
\begin{eqnarray}
   \langle T_{\rm ab}\rangle_{\rm ren}=-{g_*\over 960\pi^2 r_0^4}g_{ab},
   \label{star}
\end{eqnarray}
where $g_*$
is a dimensionless number determined by the number and spins of matter fields
existing in the de~Sitter/anti-de~Sitter space with radius $r_0$ 
($r_0=\alpha$ for anti-de~Sitter space, $r_0=\beta$ for de~Sitter space,
in practice $g_*\sim100$). Correspondingly, with the appearance of
many matter fields with de-Sitter/anti-de~Sitter invariant vacua,
Eq.~(\ref{sola}) and (\ref{solb}) should be replaced by
\begin{eqnarray}
   \Lambda_1=-{3\over\alpha^2}\left(1+{g_*\over360\pi\alpha^2}\right),
   \label{sold}
\end{eqnarray}
and
\begin{eqnarray}
   \Lambda_2={3\over\beta^2}\left(1-{g_*\over 360\pi\beta^2}\right).
   \label{sole}
\end{eqnarray}
$\Lambda_1$ is always negative. $\Lambda_2$ is negative 
if $\beta<(g_*/360\pi)^{1/2}$, positive if $\beta>(g_*/360\pi)^{1/2}$.
And, $\Lambda_1=\Lambda_2$ if 
\begin{eqnarray}
    {1\over \beta^2}-{1\over\alpha^2}=360\pi g_*^{-1}.
    \label{solc1}
\end{eqnarray}
If Eqs.~(\ref{ab}) and (\ref{solc1}) are satisfied simultaneously, the
spacetime obtained by gluing the de~Sitter space to the anti-de~Sitter
space is self-consistent and has a unique negative bare cosmological constant
through both de~Sitter and anti-de~Sitter regions. Such a solution is
very interesting. The semi-classical Einstein equations
with a negative bare cosmological constant can have two self-consistent 
solutions, one being an anti-de~Sitter space, the other being a de~Sitter
space. These two spacetimes could transit from one to the other across
a bubble wall (through
whatever quantum processes) without changing the cosmological constant.

But, for Misner-type spaces (Misner space, 
Misner-like de~Sitter space, or Misner-like anti-de~Sitter space), 
the situation for matter fields other than the massless and conformally 
coupled scalar field is more complicated since the self-consistent
vacua are not Lorentzian or de~Sitter/anti-de~Sitter invariant. 
To see this, let us consider the 
electromagnetic field in Misner space. As an alternative to 
the method of images, Euclidean
quantization is another more powerful tool for dealing with
quantum field theory in an acausal space \cite{haw79,haw95}. 
The Euclidean
method provides a convenient bridge between the conical space around a cosmic
string \cite{got85,his85} and
Misner space, which could conveniently translate the results of quantum
field theory in a conical space to that in Misner space (since a
conical space and Misner space have the same Euclidean section --- the
Euclidean conical space). In \cite{cas97},
from the well-known renormalized stress-energy tensor of the conformally
coupled scalar field in the conical space around a cosmic string, using the Euclidean
method (first translate the results in the conical space around the 
string to that in the
Euclidean conical space, then translate these results
to that in Misner space), Cassidy has successfully 
predicted that there should be a quantum state
with vanishing renormalized stress-energy tensor 
in Misner space when the boost parameter $\eta_0$ is $2\pi$, which corresponds
to no cosmic string. With Li and Gott's independent discovery of the self-consistent
vacuum state (an adapted Rindler vacuum)
for a conformally coupled scalar field in Misner space \cite{li98}, 
Cassidy's prediction has been confirmed. And, the Euclidean quantization
procedure gives a beautiful geometrical explanation 
for the self-consistent
vacua in Misner-type spaces namely that when $\eta_0=2\pi$ the corresponding
Euclidean section is flat with no conical singularity and thus has
$\langle T_\mu^{~\nu}\rangle_{\rm ren}=0$ throughout
\cite{li98,got98}. Recently, with the method of
Euclidean quantization, Li and Gott have found a self-consistent vacuum for
a model of inflation in the Kaluza-Klein theory, and found a relation
between the fine structure constant and the inflationary energy scale
which is consistent with the energy scale usually talked about in inflation
and GUT theory \cite{li98a}. Thus, we adopt the method of Euclidean 
quantization.

The Euclidean section of
Misner space is a Euclidean conical space with metric 
$ds^2=d\xi^2+\xi^2 d\phi^2+dy^2+dz^2$ where $(\xi,\phi,y,z)$ is identified
with $(\xi,\phi+n\phi_0,y,z)$ ($n=\pm1,\pm 2,...$). If we make the continuation
$\phi\rightarrow i\eta$ and $\phi_0\rightarrow i\eta_0$, we obtain Misner
space \cite{li98}. (On the other hand, if we make the continuation 
$y\rightarrow it$, we obtain the spacetime of a cosmic string.) The quantum 
field effects in the conical space 
have been investigated by many people (see \cite{dow90,iel98} and 
references therein). Thus, for the electromagnetic field in the Euclidean conical
space, the renormalized stress-energy tensor of vacuum polarization is
\begin{eqnarray}
   \langle T_\mu^{~\nu}\rangle_{\rm ren}=
   {1\over 720\pi^2\xi^4}\left[\left(2\pi\over\phi_0\right)^2-1\right]
   \left[\left(2\pi\over\phi_0\right)^2+11\right]
   \left(\begin{array}{cccc}
        1& 0& 0& 0\\
        0&-3& 0& 0\\
        0& 0& 1& 0\\
        0& 0& 0& 1
        \end{array}
        \right),
   \label{tab1}
\end{eqnarray}
where the cylindrical coordinates $(\xi,\phi,y,z)$ are used. Clearly, if
$\phi_0=2\pi$, the renormalized stress-energy tensor of the 
electromagnetic field is zero and thus the semi-classical vacuum
Einstein equations (i.e. the Einstein equations with the renormalized 
stress-energy tensor being that of vacuum polarization) are satisfied.
This is what we expect since for $\phi_0=2\pi$ the Euclidean conical
space becomes the regular flat Euclidean space $R^4$. 
But, when the Euclidean conical
space is continued to Misner space by $\phi\rightarrow i\eta$
and $\phi_0\rightarrow i\eta_0$ in Eq.~(\ref{tab1}), 
the renormalized stress-energy tensor
becomes
\begin{eqnarray}
   \langle T_\mu^{~\nu}\rangle_{\rm ren}=
   {1\over 720\pi^2\xi^4}\left[-\left(2\pi\over\eta_0\right)^2-1\right]
   \left[-\left(2\pi\over\eta_0\right)^2+11\right]
   \left(\begin{array}{cccc}
        -3& 0& 0& 0\\
        0& 1& 0& 0\\
        0& 0& 1& 0\\
        0& 0& 0& 1
        \end{array}
        \right),
   \label{tab2}
\end{eqnarray}
where the Rindler coordinates $(\eta,\xi,y,z)$ are used. We see that,
unlike the case for a massless conformally coupled scalar field, 
even if $\eta_0=2\pi$, the renormalized stress-energy tensor given by
Eq.~(\ref{tab2}) is nonzero. In fact, 
$\langle T_\mu^{~\nu}\rangle_{\rm ren}$ in Eq.~(\ref{tab2}) diverges at 
$\xi=0$
unless $\eta_0=2\pi/\sqrt{11}$. If $\eta_0=2\pi/\sqrt{11}$, the 
$\langle T_\mu^{~\nu}\rangle_{\rm ren}$ given by Eq.~(\ref{tab2})
is zero and thus it would solve the semi-classical Einstein equations
for this locally flat space. But,
if this were a self-consistent solution, it would be surprising that the
self-consistent vacua of different matter fields have different 
values of $\eta_0$.
Recall that, for the massless conformally coupled scalar field
the self-consistent vacuum has $\eta_0=2\pi$ (\cite{li98,got98}, and 
Sec.~III of this paper). And, for the case with $\eta_0=2\pi$, there is
an excellent geometric explanation: the corresponding Euclidean section
with $\phi_0=2\pi$ is the regular Euclidean space $R^4$ without conical
singularity \cite{li98,got98}; while for $\eta_0=2\pi/\sqrt{11}$, we cannot
find any simple geometric explanation. Another surprise would be that, 
for the case of the
electromagnetic field, the Euclidean
section with $\phi_0=2\pi$ is a self-consistent solution of the
semi-classical Euclidean 
Einstein equations; but, by adopting the continuation $\phi\rightarrow
i\eta$ and $\phi_0\rightarrow i\eta_0$, the resultant Misner space
with $\eta_0=2\pi$ is {\em not} 
a self-consistent solution of the semi-classical
Einstein equations. This implies that the semi-classical 
Einstein equations are broken
during this continuation: that is, with this particular
continuation the solutions of the semi-classical
Einstein equations cannot be translated between the Lorentzian and
Euclidean sections. This raises a question as to the procedure of
Euclidean quantization: is the Euclidean quantization still valid for the
electromagnetic field in Misner space? A fundamental requirement for
Euclidean quantization should be that {\em during the continuation between
the Lorentzian section and the Euclidean section, the Einstein equations
should be preserved} (otherwise the Euclidean quantization loses its
significance). Note that if we first take $\phi_0=2\pi$ in the 
Euclidean section of Misner space, $\langle T_\mu^{~\nu}\rangle_{\rm ren}$
will be zero in the Euclidean section according to 
Eq.~(\ref{tab1}). Then, if we use the continuation 
$\phi\rightarrow i\eta$ and $\phi_0$ (in this case $=2\pi$) $\rightarrow i\eta_0$ 
(in this case $=i2\pi$)
when going
from the Euclidean conical space to Misner space, naturally we expect ``zero''
should be continued to ``zero'' in the renormalized
stress-energy tensor. Then we should expect that the
$\langle T_\mu^{~\nu}\rangle_{\rm ren}$ 
in the Misner space should also be zero, which would be conflict with the results
obtained above by going to the Lorentzian section first and then setting
$\eta_0=2\pi$. What causes this non-self-consistency?

If we check the continuation procedure carefully, we find
that the problem is at renormalization. In the Euclidean section, the
renormalization is equivalent to subtracting from the original 
non-renormalized Hadamard function $G^{(1)}(X,X^\prime)$
a reference Hadamard function $G^{(1)}_{\rm ref}(X,X^\prime)$
with $\phi_0=2\pi$, i.e. the regularized Hadamard function is
\begin{eqnarray}
  G^{(1)}_{\rm reg}(X,X^\prime;\phi_0)= G^{(1)}(X,X^\prime;\phi_0)
  -G^{(1)}_{\rm ref}(X,X^\prime;2\pi),
\end{eqnarray}
where
\begin{eqnarray}
 G^{(1)}_{\rm ref}(X,X^\prime;2\pi)\equiv G^{(1)}(X,X^\prime;\phi_0=2\pi).
\end{eqnarray}
If we make the continuation with $\phi\rightarrow i\eta$ and 
$\phi_0\rightarrow i\eta_0$ but keep $\phi_0=2\pi$ unchanged in 
$G^{(1)}_{\rm ref}$, $G^{(1)}_{\rm ref}$ would be continued as
\begin{eqnarray}
 G^{(1)}_{\rm ref}(\xi,\phi,y,z;\xi^\prime,\phi^\prime,y^\prime,z^\prime;
 \phi_0=2\pi)\rightarrow
 G^{(1)}_{\rm ref}(\xi,i\eta,y,z;\xi^\prime,i\eta^\prime,y^\prime,z^\prime;
 \phi_0=2\pi=i\eta_0),
 \label{continue1}
\end{eqnarray}
then the obtained $G^{(1)}_{\rm ref}$ in the Lorentzian section is
the usual Hadamard function for the Minkowski 
vacuum (see \cite{bir82}), and the corresponding
renormalized stress-energy tensor (which is obtained by operating on 
$G^{(1)}_{\rm reg}$ with a differential operator \cite{bir82}) is
just given by Eq.~(\ref{tab2}). But, in this continuation procedure,
it would be surprising why in both $G^{(1)}_{\rm reg}$ and $G^{(1)}$
the parameter $\phi_0$ was changed to $i\eta_0$ but in $G^{(1)}_{\rm ref}$
the parameter $\phi_0=2\pi$ was unchanged, thus this procedure is not
self-consistent. The result of this non-self-consistent procedure is
that the semi-classical Einstein equations are broken, as mentioned
above. 

For a flat Euclidean space $R^4$ with Cartesian coordinates 
$(\tau,x,y,z)$ and metric $ds^2=d\tau^2+dx^2+dy^2+dz^2$, if we go to the
Lorentzian section by the continuation $\tau\rightarrow it$, the
Euclidean space $R^4$ is naturally continued to the usual simply connected
Minkowski space with Cartesian coordinates $(t,x,y,z)$ and metric 
$ds^2=-dt^2+dx^2+dy^2+dz^2$. Since in the Euclidean section $\tau$ goes
from $-\infty$ to $\infty$, naturally in the Lorentzian section $t$
also goes from $-\infty$ to $\infty$. With this continuation, the
Hadamard function for the Euclidean vacuum in the Euclidean $R^4$ space 
--- which is the reference Hadamard function for renormalization in the
Euclidean section --- is
continued to the Hadamard function for the Lorentzian vacuum in Minkowski space
--- which is the reference Hadamard function for renormalization in 
Minkowski space. But in our case, we start with a flat Euclidean space
$R^4$ with cylindrical coordinates $(\xi,\phi,y,z)$ and metric
$ds^2 = d\xi^2 +\xi^2 d\phi^2 +dy^2 +dz^2$ where $\phi$ has a period of
$2\pi$. When we go to the Lorentzian section by the continuation
$\phi\rightarrow i\eta$, very naturally $\eta$ has also a period of
$2\pi$. Then the obtained Lorentzian section is a Misner space with
the boost parameter $\eta_0=2\pi$. With this continuation, the Hadamard
function for the Euclidean vacuum in the Euclidean section ---  
which is the reference Hadamard function for renormalization in the
Euclidean section --- is continued to the Hadamard function
for Li and Gott's adapted Rindler vacuum with $\eta_0=2\pi$
--- very naturally which should also be the reference Hadamard function
for renormalization in Misner space. We start with the reference space
for a Euclidean conical space --- which is the flat Euclidean space
without conical singularity, and end with the corresponding the reference space for
Misner space --- which is the Misner space with $\eta_0=2\pi$. This
procedure is very natural. Therefore, here we propose
the following {\em self-consistent renormalization procedure} 
for quantum fields in Misner-type spaces: 
When we make the continuation from the Euclidean section to the
Lorentzian section by $\phi\rightarrow i\eta$ and $\phi_0\rightarrow 
i\eta_0$, we should also make the continuation $2\pi\rightarrow i2\pi$
(see Fig.~\ref{fig6}), then the reference Hadamard function is
continued to be the Hadamard function in the Lorentzian section
(Misner space) with $\eta_0=2\pi$. 
With this self-consistent renormalization procedure, instead of 
Eq.~(\ref{continue1}), the reference Hadamard function is continued
to be
\begin{eqnarray}
 G^{(1)}_{\rm ref}(\xi,\phi,y,z;\xi^\prime,\phi^\prime,y^\prime,z^\prime;
 \phi_0=2\pi)\rightarrow
 G^{(1)}_{\rm ref}(\xi,i\eta,y,z;\xi^\prime,i\eta^\prime,y^\prime,z^\prime;
 \eta_0=2\pi),
 \label{continue2}
\end{eqnarray}
making all the problems mentioned above go away.
With this self-consistent renormalization procedure, 
instead of Eq.~(\ref{tab2}), the renormalized stress-energy 
tensor of electromagnetic field in Misner space becomes
(substituting $i\eta_0$ for $\phi_0$ and
$i2\pi$ for $2\pi$ in Eq.~(\ref{tab1}))
\begin{eqnarray}
   \langle T_\mu^{~\nu}\rangle_{\rm ren}=
   {1\over 720\pi^2\xi^4}\left[\left(2\pi\over\eta_0\right)^2-1\right]
   \left[\left(2\pi\over\eta_0\right)^2+11\right]
   \left(\begin{array}{cccc}
        -3& 0& 0& 0\\
        0& 1& 0& 0\\
        0& 0& 1& 0\\
        0& 0& 0& 1
        \end{array}
        \right),
   \label{tab3}
\end{eqnarray}
in the Rindler coordinates $(\eta,\xi,y,z)$. Clearly, if $\eta_0=2\pi$, we 
have $\langle T_\mu^{~\nu}\rangle_{\rm ren}=0$ and thus the 
semi-classical Einstein equations are satisfied. The situation is similar
for massless neutrinos in Misner space. If we make the continuation $\phi
\rightarrow i\eta$, $\phi_0\rightarrow i\eta_0$, $2\pi\rightarrow i2\pi$,
using the self-consistent
renormalization procedure, we obtain the renormalized stress-energy
tensor of vacuum polarization for massless neutrinos in Misner space
\begin{eqnarray}
   \langle T_\mu^{~\nu}\rangle_{\rm ren}=
   {1\over 5760\pi^2\xi^4}\left[\left(2\pi\over\eta_0\right)^2-1\right]
   \left[\left(2\pi\over\eta_0\right)^2+17\right]
   \left(\begin{array}{cccc}
        -3& 0& 0& 0\\
        0& 1& 0& 0\\
        0& 0& 1& 0\\
        0& 0& 0& 1
        \end{array}
        \right),
   \label{tab4}
\end{eqnarray}
which is also zero for $\eta_0=2\pi$. Thus, with the new renormalization
procedure, we obtain self-consistent vacua for electromagnetic fields
and massless neutrinos in Misner
space. It is very easy to check that this new renormalization procedure
does not change the results for a massless conformally coupled
scalar field in Misner space. The Euclidean result for a conical
space is
\begin{eqnarray}
   \langle T_\mu^{~\nu}\rangle_{\rm R,ren}={1\over1440\pi^2\xi^4}
   \left[\left({2\pi\over \phi_0}\right)^4-1\right]
   \left(\begin{array}{cccc}
   -3&0&0&0\\
   0&1&0&0\\
   0&0&1&0\\
   0&0&0&1
   \end{array}
   \right),
   \label{E69a}
\end{eqnarray} 
Substituting $i\eta_0$ for $\phi_0$ and $i2\pi$ for $2\pi$ into
Eq.~(\ref{E69a}) gives Eq.~(\ref{E69}) just as before. Thus,
Li and Gott's adapted Rindler vacuum \cite{li98} is still a self-consistent
vacuum for the massless conformally coupled scalar field in
Misner space with $\eta_0=2\pi$.
These results can be easily transplanted to Misner-like de~Sitter
space and Misner-like anti-de~Sitter space, since these fields (massless
and conformally coupled scalar fields, electromagnetic fields, and
massless neutrinos) are
conformally invariant and de~Sitter space and anti-de~Sitter space are
conformally flat. The results are: if the boost period is $2\pi r_0$
(where $r_0$ is the radius of the de~Sitter space or anti-de~Sitter space),
the renormalized stress-energy tensors are finite and given by 
Eq.~(\ref{star}) with $g_*=11/2$ for one neutrino field and $g_*=62$ for
the electromagnetic field. With Eqs.~(\ref{sold}) and (\ref{sole}), 
self-consistent solutions of the semiclassical Einstein equations with
cosmological constant can be found.

Thus, with our new renormalization method, various self-consistent 
vacuum states in Misner-type spaces with CTCs can be found.

The chronology protection conjecture \cite{haw92} was originally based on the
fact that for a massless and conformally coupled scalar field with the
adapted Minkowski vacuum in the Misner space the renormalized 
stress-energy tensor of vacuum polarization
diverges  at the chronology horizon \cite{his82}. After
the appearance of many counter-examples 
\cite{bou92,li93,li94,tan95,li96,kra96,sus97,vis97,cas97}, Cassidy and
Hawking \cite{cas98} demonstrated that the back-reaction of vacuum
polarization does not enforce chronology protection. The results in this
paper (and \cite{li98,got98}) support this demonstration. However,
Cassidy and Hawking \cite{cas98} turned to the proposition that
the effective action of matter fields in spacetimes with CTCs always diverges
at the chronology horizon and that it is this that  enforces chronology protection.
The effective action plays an important role in Euclidean quantum gravity,
which gives the probability for the creation of spacetime
through quantum tunneling. Consider a
simple example of a massless conformally coupled scalar field in
Misner space. The Euclidean effective Lagrangian density is \cite{iel98}
\begin{eqnarray}
   {\cal L}={1\over 1440\pi^2\xi^4}\left[\left(2\pi\over\phi_0\right)^2
   -1\right]\left[\left(2\pi\over\phi_0\right)^2+11\right].
   \label{lag}
\end{eqnarray}
The Euclidean effective action is obtained by integrating the Euclidean
effective Lagrangian density over a suitable volume of the Euclidean section. 
Clearly, if $\phi_0\neq 2\pi$, the effective Euclidean
Lagrangian diverges at $\xi=0$ (thus the Euclidean effective action 
diverges if the domain for integration includes the conical singularity at
$\xi=0$). But, if $\phi_0=2\pi$, we have ${\cal L}=0$, which is not surprising
since when $\phi_0=2\pi$ the conical singularity at $\xi=0$ disappears
and the Euclidean section becomes the usual regular flat $R^4$ space. The
Euclidean effective action is also zero since the Euclidean effective 
Lagrangian is zero everywhere. By the continuation $\phi\rightarrow
i\eta$, $\phi_0\rightarrow i\eta_0$, and with the self-consistent
renormalization procedure outlined above, the reference Hadamard function
is continued to be that with $\eta_0=2\pi$ 
in Misner space (which means that the ratio
$2\pi/\phi_0$ in Eq.~(\ref{lag})
is changed to $2\pi/\eta_0$ since $\phi_0\rightarrow i\eta_0$
and $2\pi\rightarrow i2\pi$), then the effective Lagrangian for a
massless conformally coupled scalar field in Misner space is
\begin{eqnarray}
   {\cal L}={1\over 1440\pi^2\xi^4}\left[\left(2\pi\over\eta_0\right)^2
   -1\right]\left[\left(2\pi\over\eta_0\right)^2+11\right],
   \label{lag1}
\end{eqnarray}
which is zero everywhere if $\eta_0=2\pi$. Thus the Lorentzian 
effective action is
also zero for $\eta_0=2\pi$. This shows that, for the self-consistent
vacuum in Misner space, the effective action is zero. It can be expected
that this result can be extended to other quantum fields or 
other spacetimes with CTCs, where
the (Euclidean or Lorentzian) effective Lagrangian would be finite throughout
the space [and thus the (Euclidean or Lorentzian) effective action would
also be finite] for the self-consistent vacua. Thus 
Cassidy and Hawking's argument that the effective action (or 
equivalently the entropy)
enforce chronology protection is questionable.  

\section{Conclusions}
From the covering space of anti-de~Sitter space, a Misner-like
anti-de~Sitter space can be constructed. 
This Misner-like anti-de~Sitter space has CTCs but the 
regions with CTCs are separated from the regions without CTCs by
chronology horizons. In the appropriate coordinates, this Misner-like
anti-de~Sitter space is just the Lorentzian section of the complex
space with CTCs constructed by Li, Xu, and Liu in 1993 \cite{li93}.
For a massless conformally coupled scalar field in this space, a
self-consistent vacuum is found, whose renormalized stress-energy
tensor is like that of a positive cosmological constant ---
which when added to an appropriate negative bare cosmological
constant can self-consistently solve the semi-classical
Einstein equations.

By gluing a de~Sitter space to an anti-de~Sitter space along a  
bubble wall, another new spacetime with CTCs is obtained. 
This spacetime could describe the transition between de~Sitter space
and anti-de~Sitter space. In this spacetime, the region with CTCs and 
the region without CTCs are separated via chronology horizons.  
For the de~Sitter/anti-de~Sitter 
invariant vacua in
these spacetimes, the renormalized stress-energy tensors are like positive
cosmological constants. A self-consistent solution can be obtained if there is
a single negative bare cosmological constant in the two regions with
the renormalized stress-energy tensor of vacuum polarization adding different
positive cosmological constants to the two sides of the bubble wall
so that the effective cosmological constant (bare $+$ renormalized) is positive
on one side and negative on the other. 
On the hypersurface separating de~Sitter space from
anti-de~Sitter space, in order that the Einstein equations are satisfied,
a surface stress-energy tensor should be induced. If the hypersurface is
timelike, the surface stress-energy tensor is like that of a three-dimensional
positive cosmological constant. 

The self-consistent solutions of the semi-classical Einstein equations
with cosmological constant and the renormalized stress-energy tensor
of vacuum polarization in de~Sitter/anti-de~Sitter space are investigated.
If the bare cosmological constant is positive, there are two self-consistent 
solutions, both of them are de~Sitter spaces. If the bare cosmological constant
is zero then there are two self-consistent solutions --- one is Minkowski space
and the other is a Planckian scale de~Sitter space. If the bare cosmological
constant is negative, there are also two self-consistent solutions,
one of them is an anti-de~Sitter space, but the other is a 
sub-Planckian scale de~Sitter space.
And, at the sub-Planckian scale, self-consistent solutions (either
de~Sitter space
or anti-de~Sitter space) exist only for a bare negative cosmological constant.

The generalization to electromagnetic fields and massless neutrinos 
in spacetimes
with CTCs are discussed. It is argued that, for Misner-type spacetimes,
in order that the semi-classical
Einstein equations are preserved under continuation between the Euclidean
and Lorentzian sections, a new renormalization procedure should be
introduced. We have proposed such a self-consistent renormalization procedure,
with which self-consistent vacua for electromagnetic fields and neutrinos
are found. 

\acknowledgments
I am very grateful to J. Richard Gott for many stimulating and
helpful discussions.
This research was supported by NSF grant AST95-29120 and NASA grant NAG5-2759.

\begin{figure}
\caption{The Penrose diagram of the covering space of 
anti-de~Sitter space. The left vertical line represents the hypersurface
in anti-de~Sitter space with $X=0$, where the global static coordinates
$(t,\chi,\theta,\phi)$ defined by Eq.~(\ref{coord1}) are singular ($\chi=0$).
The right vertical line labeled with ${\cal J}$ represents null infinity 
($\chi=\infty$). The horizontal dashed lines represent 
hypersurfaces with $t={\rm constant}$, the labels $0$, $\pi$, and $2\pi$
refer $t/\alpha=0$, $\pi$, and $2\pi$ respectively. The grey triangle
represents the region covered by the non-stationary (open cosmological)
coordinates 
$(t,\chi,\theta,\phi)$ defined by Eqs.~(\ref{coord3}) and 
(\ref{metr3}). The two isolated points
labeled with $i^+$ and $i^-$ represent future timelike infinity
and past timelike infinity respectively. If the 
hypersurfaces with the global time $t/\alpha=0$ and $t/\alpha=2\pi$ are
identified, we obtain the usual anti-de~Sitter space with CTCs everywhere.}
\label{fig1}
\end{figure}

\begin{figure}
\caption{The Penrose diagrams of the Misner-like anti-de~Sitter space
constructed by identifying points related by boost transformations
in the covering space of anti-de~Sitter space. With these boost
transformations, $A$ and $A^\prime$ are unchanged.
(a) In the left diagram, the light and dark
grey regions represent unit cells in the 
Misner-like anti-de~Sitter space, whose opposite boundaries
(heavy dashed lines) being identified. The chronology horizons ${\cal CH}^+$
and ${\cal CH}^-$ separate the regions with CTCs (the dark grey regions) 
from that without CTCs (the light grey region).
(b) The right diagram is equivalent to the left one, except that the 
fundamental cell is chosen to be one bounded with null hypersurfaces.
$E^\prime$ is the image of $E$ under the boost transformation.}
\label{fig2}
\end{figure}

\begin{figure}
\caption{The Penrose diagram of de~Sitter space. The event $E^\prime$ is the 
anti-podal point of the event $E$. The upper horizontal
line labeled with ${\cal J}^+$ represents future null infinity, the lower
horizontal line labeled with with ${\cal J}^-$ represents
past null infinity. Two (future and past) light cones from $E$ and $E^\prime$
are shown. The curve labeled with $\Sigma_1$ represents a timelike hypersurface
with $W={\rm constant}<\beta$. The curves labeled with $\Sigma_2^{~\pm}$
represent the spacelike hypersurfaces with $W={\rm constant}>\beta$
(two leaves).}
\label{fig3}
\end{figure}

\begin{figure}
\caption{A schematic diagram for the spacetime obtained by gluing a de~Sitter
space to an anti-de~Sitter space along a timelike wall. The vertical hyperbola
of one sheet (to the left)
represents a de~Sitter space in the embedding space of Eq.~(\ref{des1}), 
the horizontal hyperbola of one sheet 
(to the right) represents 
an anti-de~Sitter space in the embedding space of Eq.~(\ref{metr0}). 
They are glued along a timelike hypersurface (a bubble wall) on which a  
surface stress-energy tensor is induced so that Einstein equations are
satisfied there. The two embedding spaces match at the bubble
wall ($W={\rm constant}$). The Penrose diagram of this spacetime is shown in 
Fig.~5.}
\label{fig4}
\end{figure}

\begin{figure}
\caption{The Penrose diagram of the spacetime obtained by gluing 
anti-de~Sitter space to de~Sitter space along a timelike hypersurface. The heavy
curve labeled with $\Sigma$ represents a timelike hypersurface(the bubble wall)
on the left side of which there is a 
de~Sitter space ($dS^4$), and on the right side of which there is an
anti-de~Sitter
space ($AdS^4$). On the anti-de~Sitter side, the spacelike hypersurfaces
denoted with two dashed lines are identified. This spacetime contains CTCs
in the grey region, but no CTCs in the blank region. The region with CTCs 
is separated from that without CTCs by the chronology horizons 
${\cal CH}^\pm$. ${\cal J}^+$, ${\cal J}^-$, and ${\cal J}$
represent future null infinity, past null infinity, and
null infinity respectively. (Compare with Fig.~\ref{fig4}.)}
\label{fig5}
\end{figure}

\begin{figure}
\caption{A schematic diagram of the old and the new renormalization procedures.
The horizontal lines are real axes, the vertical lines are imaginary axes. The arcs
with arrows represent the continuation from the Euclidean section to the
Lorentzian section.
(a) The left diagram describes the old renormalization procedure in Euclidean
quantization. As one goes from the Euclidean section of a conical space to its
Lorentzian section (Misner space), $\phi$ is changed to $i\eta$, $\phi_0$ is
changed to $i\eta_0$, while the parameter $2\pi$ in the 
reference Hadamard function is unchanged. With this old renormalization procedure,
the semi-classical
Einstein equations are broken during the continuation, as discussed in the
text. (b) The right diagram describes the new self-consistent renormalization
procedure in Euclidean quantization. With this new procedure, as one goes from
the Euclidean section to the Lorentzian section, $\phi$ is changed to $i\eta$, 
$\phi_0$ is changed to $i\eta_0$, and $2\pi$ in the reference Hadamard 
function is changed to $i2\pi$. The semi-classical
Einstein equations are preserved during this new continuation.}
\label{fig6}
\end{figure}

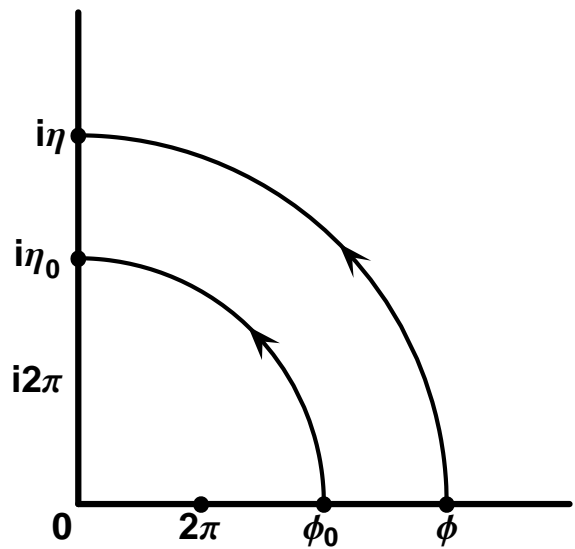
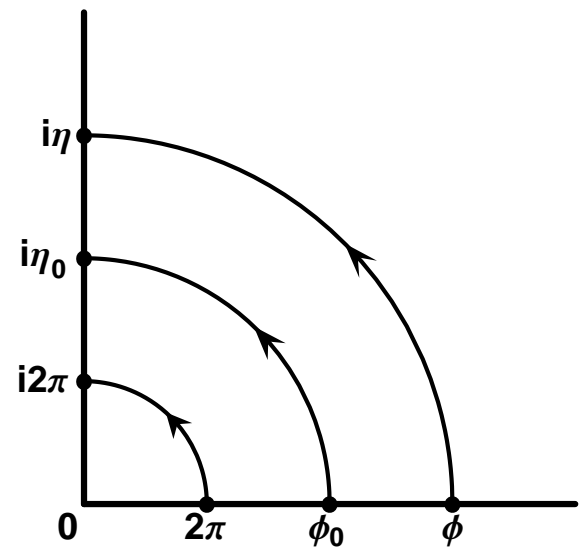
\end{document}